\documentclass[reprint,aps,prl,superscriptaddress]{revtex4-2}
\usepackage{graphicx}
\usepackage{bm}
\usepackage{bbold}
\usepackage{CJK}
\usepackage{amsthm}
\usepackage{multirow}
\usepackage{xcolor}
\usepackage{threeparttable}

\usepackage{amsmath}
\DeclareMathOperator{\Tr}{Tr}
\DeclareMathOperator{\A}{\mathcal{A}}

\newcommand{\ket}[1]{ |{#1} \rangle}

\theoremstyle{plain}
\newtheorem*{theorem*}{Theorem}

\begin{document}

\title{Experimental Examination of Entanglement Estimates}
\author{Songbo Xie}
\affiliation{Center for Coherence and Quantum Optics, and Department of Physics and Astronomy, University of Rochester, Rochester, New York 14627, USA}

\author{Yuan-Yuan Zhao}
\email{zhaoyy01@pcl.ac.cn}
\affiliation{Peng Cheng Laboratory, Shenzhen 518052 People's Republic of China}

\author{Chao Zhang}
\affiliation{CAS Key Laboratory of Quantum Information, University of Science and Technology of China, Hefei 230026, People’s Republic of China}
\affiliation{CAS Center For Excellence in Quantum Information and Quantum Physics, University of Science and Technology of China, Hefei 230026, People’s Republic of China}
\affiliation{Hefei National Laboratory, University of Science and Technology of China, Hefei 230088, China}

\author{Yun-Feng Huang}
\affiliation{CAS Key Laboratory of Quantum Information, University of Science and Technology of China, Hefei 230026, People’s Republic of China}
\affiliation{CAS Center For Excellence in Quantum Information and Quantum Physics, University of Science and Technology of China, Hefei 230026, People’s Republic of China}
\affiliation{Hefei National Laboratory, University of Science and Technology of China, Hefei 230088, China}

\author{Chuan-Feng Li}
\affiliation{CAS Key Laboratory of Quantum Information, University of Science and Technology of China, Hefei 230026, People’s Republic of China}
\affiliation{CAS Center For Excellence in Quantum Information and Quantum Physics, University of Science and Technology of China, Hefei 230026, People’s Republic of China}
\affiliation{Hefei National Laboratory, University of Science and Technology of China, Hefei 230088, China}

\author{Guang-Can Guo}
\affiliation{CAS Key Laboratory of Quantum Information, University of Science and Technology of China, Hefei 230026, People’s Republic of China}
\affiliation{CAS Center For Excellence in Quantum Information and Quantum Physics, University of Science and Technology of China, Hefei 230026, People’s Republic of China}
\affiliation{Hefei National Laboratory, University of Science and Technology of China, Hefei 230088, China}

\author{Joseph H. Eberly}
\affiliation{Center for Coherence and Quantum Optics, and Department of Physics and Astronomy, University of Rochester, Rochester, New York 14627, USA}

\date{\today}

\begin{abstract}
Recently a proper genuine multipartite entanglement (GME) measure has been found for three-qubit pure states [see Xie and Eberly, Phys. Rev. Lett. {\bf 127}, 040403 (2021)], but capturing useful entanglement measures for mixed states has remained an open challenge. So far, it requires not only a full tomography in experiments, but also huge calculational labor. A leading proposal was made  by G\"uhne, Reimpell, and Werner [Phys. Rev. Lett. \textbf{98}, 110502 (2007)], who used expectation values of entanglement witnesses to describe a lower bound estimation of entanglement. We provide here an extension that also gives genuine upper bounds of entanglement. This advance requires only the expectation value of {\em any} Hermitian operator. Moreover, we identify a class of operators $\A_1$ which not only give good estimates, but also require a remarkably small number of experimental measurements. In this note we define our approach and illustrate it by estimating entanglement measures for a number of pure and mixed states prepared in our recent experiments.
\end{abstract}

\maketitle

{\it Introduction.}---The concept of entanglement can be regarded as a surprising ``side effect'' of the long-lasting Bohr-Einstein debates on quantum mechanics. Initially, entanglement was only identified as the non-factorizability of quantum states for two or more parties. Progress in quantum information and quantum communication (examples are \cite{ekert1991,bennett1992,bennett1993,bennett1996}) exposed the practical usefulness of entanglement as well as the need for quantification of entanglement as a resource in various tasks. The solution to the quantification problem was first developed as an entropy for pure states \cite{von1955}. The focus for further work has moved to what is now called a ``genuine entanglement measure" (GME - see \cite{ma2011}). 

The significant participation of mixed states in practical experiments puts attention on the absence of any convenient GME measures for mixed states. Among the possibilities to address the well known difficulty of defining and evaluating mixed-state measures, an attractive one is based on convex-roof construction \cite{uhlmann1998}. For this one begins with a pure-state entanglement measure $E(\psi)$. The extension to all mixed states is given by
\begin{equation}\label{convexroof}
    E(\rho) = \min_{\{p_i,\psi_i\}}\sum_ip_iE(\psi_i).
\end{equation}
The optimization is taken over all possible de-compositions of $\rho=\sum_ip_i|\psi_i\rangle\langle\psi_i|$. A mixed state $\rho$ contains no genuine entanglement if and only if $E(\rho)=0$.

An analytic expression of \eqref{convexroof} for two-qubit systems was successfully derived for the concurrence measure \cite{hill1997}, but a systematic evaluation for more complicated systems has been challenging and is difficult even numerically \cite{rothlisberger2009}, not to mention the separate difficulty of making a faithful measurement of mixed-state entanglement in a real experiment.

In \cite{audenaert2006}, Audenaert and Plenio took a key step forward by exploring the conditions when correlations, or other measurement data are able to reveal entanglement of the system. Later developments (see \cite{guhne2007,eisert2007}) advocated the use of {\it entanglement witness operators} with the help of Legendre transformations to give lower-bound (LB) estimations to the expression \eqref{convexroof} for an arbitrary entanglement measure. These works provide, in principle, a viable method to estimate entanglement by a lower bound in real experiments.

However, we note that upper-bound (UB) estimates are still missing. While a nonzero LB estimate detects the existence of entanglement, a vanishing result of LB estimation does not guarantee separability. In order to have a quantifiable idea of entanglement for a state prepared in a real experiment, both LB and UB are needed. In fact, only when the UB estimate is 0 can one be confident that the quantum state is separable. In some situations, an exact confirmation of no entanglement is important, a prominent example being in studies of the sudden death of entanglement \cite{yu2009}.

A few previous attempts for entanglement upper bounds were limited to specific entanglement measures (for example, the {\it entanglement of formation} in \cite{zhu2012} and the {\it R\'enyi-$\alpha$ entropy} in \cite{song2016}). These usually cannot be extended to more complicated systems, multipartite entanglement being a key example. On the other hand, by itself the definition in Eq. \eqref{convexroof} implies upper bounds of $E(\rho)$, but the search for a decomposition $\{p_i,\psi_i\}$ requires a full tomography of $\rho$. The advance we advocate here requires an entirely new motivation: we are describing a simple way to estimate an UB of entanglement requiring remarkably few measurements in the experiment.

With this change in focus, we have developed a mathematical structure to reinterpret the results in \cite{guhne2007}. Our new structure has the advantage to be easily extended to an upper-bound estimation of entanglement. We show below that without doing a full tomography of quantum states in a real experiment, the measurement of a remarkably small number of Hermitian bases is quite enough to provide information for both lower bounds and upper bounds universally, that is, for {\it any} entanglement measure. We report an examination of this claim in a specific experiment. 

{\it Lower-Bound and Upper-Bound Estimators.}---In quantum physics, the space of Hermitian operators for $n-$qubit system has $4^n$ dimensions. A convenient basis set for experimental measurement is the ``Pauli products'' $\{\sigma_i\otimes\sigma_j\otimes\cdots\}$, with $\{i,j,\cdots\}=\{0,x,y,z\}$, where $\sigma_0=\mathbb{1}$, and $\sigma_{x},\sigma_y,\sigma_z$ are the usual Pauli matrices.

Among all the Hermitian matrices, there is a special subset of operators called {\it Tight Lower-Bound Estimators}. The expectation values of these operators, when given an arbitrary entanglement measure $E$, {\it do not overestimate} the entanglement values for {\it any} mixed states.

{\it Definition 1}.---A Hermitian operator $X$ is called a {\it Tight Lower-Bound Estimator} if and only if it satisfies the two conditions: (a)\ $\Tr(X\rho)\leq E(\rho),\ \forall\ \text{mixed state } \rho$. (b) $\exists\ \text{mixed state }\sigma$ such that $\Tr(X\sigma)=E(\sigma)$.
    
The first condition ensures that the expectation value of $X$ does not overestimate the measure $E$ for any mixed state $\rho$. This is far from enough, since any operator of the form $\A-\lambda\mathbb{1}$, where $\lambda$ is an extremely large number, satisfies condition (a). However, these extremely ``low'' operators (the expectation value being extremely negative) cannot provide a useful lower bound for entanglement. The second condition fixes the problem by ensuring ``tightness'', that is, the value $(-\lambda)$ shall be chosen as large as possible. 

\begin{figure}[t]
    \centering
    \includegraphics[width=0.4\textwidth]{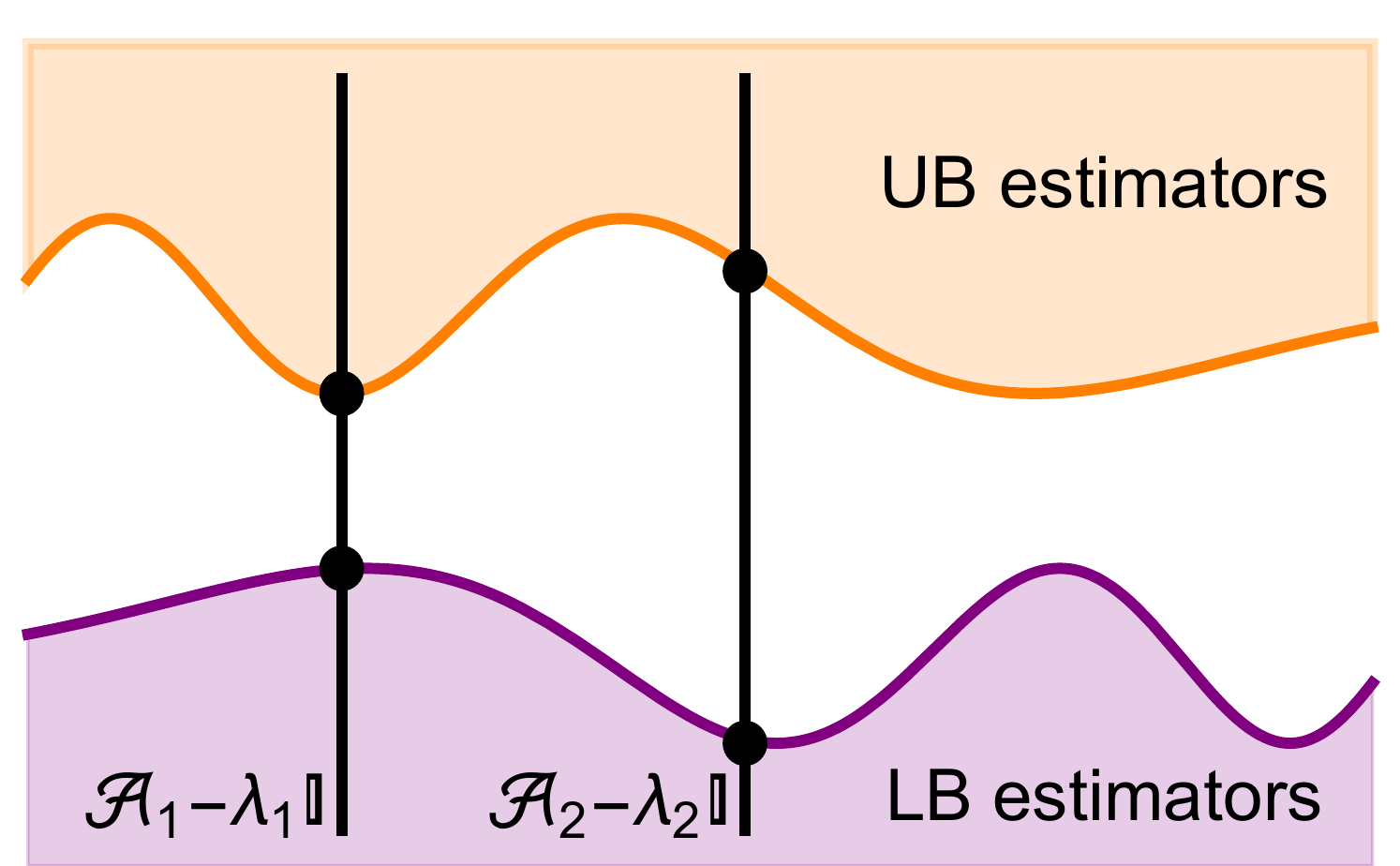}
    \caption{
    A simple illustration of the fiber bundle structure in the 64-dimensional Hermitian operator $\mathcal{A}$ space. The black vertical lines represent the two fibers $\mathcal{A}_1-\lambda_1\mathbb{1}$ and  $\mathcal{A}_2-\lambda_2\mathbb{1}$. The horizontal axis is a heuristic illustration of the 63-dimensional subspace perpendicular to the fibers. The purple region represents all the LB estimators, that is, these operators do not overestimate the entanglement for all mixed states. However, only the operators on the top boundary (in dark purple color, and they form a section of the fiber bundle) satisfy the ``tight'' property, as is explained in the text. Similarly, the orange region contains all the UB estimators, while only the operators on the bottom boundary (in dark orange color, and they form another section) are ``tight''. In the figure, one can tell the fiber $\mathcal{A}_1-\lambda_1\mathbb{1}$ provides better entanglement estimates, since its UB and LB are closer to each other than the fiber $\mathcal{A}_2-\lambda_2\mathbb{1}$.
    }
    \label{fig:fiber}
\end{figure}

From the results in \cite{guhne2007,eisert2007}, we notice that it is equivalent to define tight LB estimators in a simpler way: (a) $\langle\psi|X|\psi\rangle\leq E(\psi),\ \forall\ \text{pure state }\psi$. (b) $\exists\ \text{pure state }\phi$ such that $\langle\phi|X|\phi\rangle=E(\phi)$. Only pure states are engaged.

In the space of Hermitian operators, we consider the one-dimensional parallel fibers $\A-\lambda\mathbb{1}$, with $\lambda$ a real number and $\mathbb{1}$ the identity matrix. That is, two operators are on the same fiber if and only if they differ by $\lambda\mathbb{1}$. The result in \cite{ryu2012} provides a proof that all the tight LB estimators form a section of the fiber bundle. That is, there exists one and only one tight LB estimator on each fiber $\A-\lambda\mathbb{1}$, as is given by $X_{\A}=\A-\lambda_\text{LB}\mathbb{1}$, with $\lambda_\text{LB}=\max_\psi\left(\langle\psi|\A|\psi\rangle-E(\psi)\right)$. The optimization is taken over all pure states $\psi$.

Up to now, only lower bounds are engaged. We show that our fiber bundle structure is able to extend these results to upper-bound estimations of entanglement.

{\it Definition 2.}---A Hermitian operator $Y$ is called a {\it Tight Upper-Bound Estimator} if and only if it satisfies the two conditions: (a) $\Tr(Y\rho)\geq E(\rho),\ \forall$ mixed state $\rho$. (b) $\exists$ mixed state $\sigma$ such that $\Tr(Y\sigma)=E(\sigma)$. An equivalent but simpler form of definition is again available with pure states engaged only: (a) $\langle\psi|Y|\psi\rangle\geq E(\psi),\ \forall\ \text{pure state }\psi$. (b) $\exists\ \text{pure state }\phi$ such that $\langle\phi|Y|\phi\rangle=E(\phi)$. 

It is a simple exercise to observe that all the tight UB estimators form another section of the fiber bundle. Specifically, there exists one and only one tight UB estimator on the fiber $\A-\lambda\mathbb{1}$, as is given by $Y_{\A}=\A-\lambda_\text{UB}\mathbb{1}$, with $\lambda_\text{UB}=\min_\psi\left(\langle\psi|\A|\psi\rangle-E(\psi)\right)$. The optimization is again taken over all pure states $\psi$.

We now show the main result: We can estimate the entanglement of an arbitrary mixed state $E(\rho)$ by measuring the expectation value of any Hermitian operator: $\langle \A\rangle_\rho-\lambda_\text{LB}\leq E(\rho)\leq\langle \A\rangle_\rho-\lambda_\text{UB}$. The expectation value $\langle \A\rangle_\rho$ is measured in experiments, while the two real numbers $\lambda_\text{LB}$ and $\lambda_\text{UB}$ are evaluated independently as described above once $\A$ is determined. A geometric structure is illustrated heuristically in Fig. \ref{fig:fiber}.

\begin{figure}[t]
	\centering
	\includegraphics[width=0.98\linewidth]{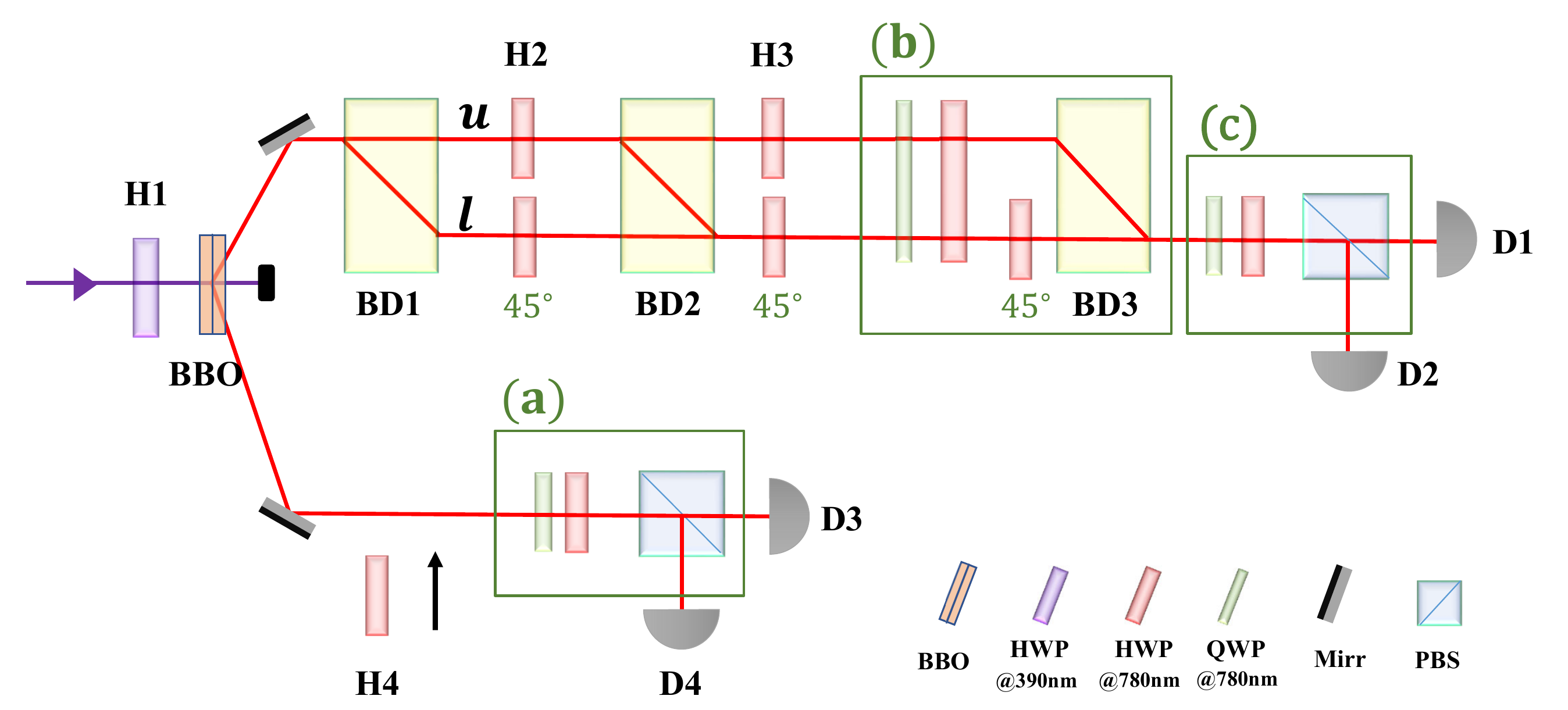}
	\caption{{\bf Experimental setup.} The experimental setup generates the three-qubit entangled states, where the configuration in boxes (a), (b), and (c) are used to perform the measurements of the single-qubit observables. Abbreviations of the components: BBO, $\beta$-barium-borate crystal; HWP, half-wave plate; QWP, quarter-wave plate; PBS, polarizing beam splitter; BD, beam splitter; D1-D4, single-photon detector. More details are explained in the text. }
	\label{fig:setup}
\end{figure}

{\it Experimental Setup.}---A sketch of the experimental setup is depicted in Fig.~\ref{fig:setup}. The methods are similar to the ones in Refs. \cite{Osetup2012,Osetup2015,Osetup2021}. A laser with a wavelength of 390~nm, a repetition rate of 80~MHz, and a pulse duration of 140~fs, pumps two crossed-axis type-I $\beta$-barium borate (BBO) crystals to generate polarization-entangled photon pairs via the spontaneous parametric down-conversion (SPDC) process. To be specific, the entangled photons are in the state $\ket{\phi}=\alpha\ket{H}_s\ket{H}_i+\beta \ket{V}_s\ket{V}_i$, where the subscripts $\{s,i\}$ represent the \{signal, idler\} photon, and $H$($V$) represent the photon as horizontally(vertically) polarized. The ratio of the parts $\ket{HH}$ and $\ket{VV}$ are changed by rotating the half-wave plate H1 with $\alpha=\cos2\theta_1$ and $\beta=\sin2\theta_1$.  

\begin{figure}[b]
    \centering
    \includegraphics[width=0.48\textwidth]{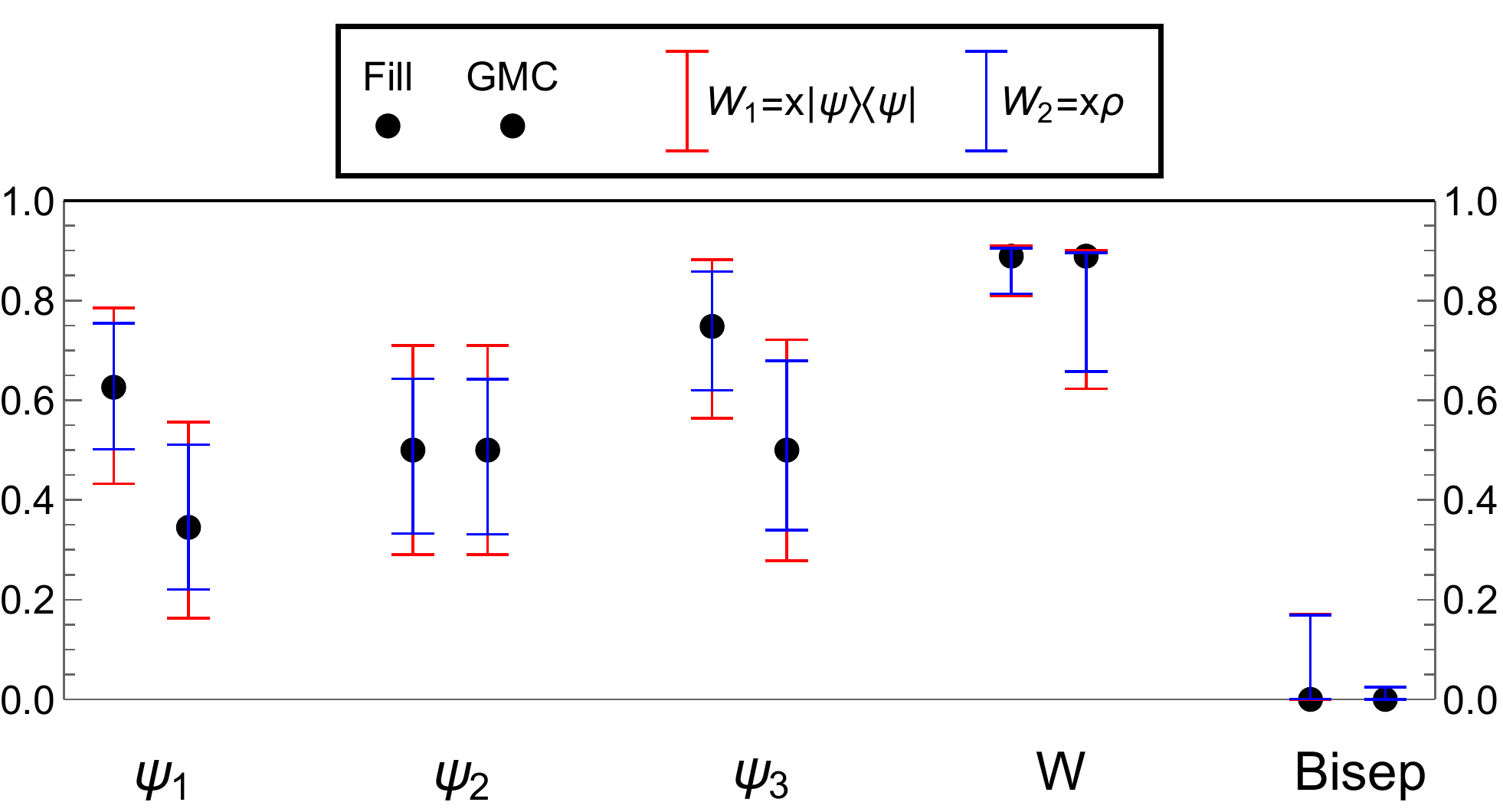}
    \caption{The entanglement bound estimations for five pure quantum states $|\psi_i\rangle$ given in \eqref{fourstate} by two entanglement measures Fill and GMC. The results are divided into five groups, each group for one state. Within each group, the left result is for the Fill measure, while the right result is for the GMC measure. For each result, the black dot represents the theoretical entanglement for the pure state $|\psi_i\rangle$, which is close to the exact entanglement in the experiment due to the high fidelity in Table \ref{tab:exp}. The red bar is the bound by using operator $\A_1$, and the blue bar is the bound by using operator $\A_2$. See details in the text.}
    \label{fig:pureresult}
\end{figure}

Various three-qubit states are then prepared by introducing the path degrees of freedom (DoFs), $u$ and $l$. That is, two of the three qubits are encoded on the polarization DoFs of the photons and the third qubit is encoded on the path DoF of the idler photon. As shown in Fig.~\ref{fig:setup}, the vertically polarized component of the idler photon passes through the beam displacer BD1, while the horizontal component is deflected with a $4~mm$ deviation. The state is now $\ket{\phi}=\alpha\ket{H}_s\ket{H}_i\ket{l}_i+\beta \ket{V}_s\ket{V}_i\ket{u}_i$, and evolves as follows:
\begin{widetext}
\begin{equation}
\begin{split}
\ket{\phi}& \xrightarrow{\text{H2},\text{BD2}} \alpha\ket{H}_s\ket{V}_i\ket{l}_i+\beta \sin2\theta_2\ket{V}_s\ket{H}_i\ket{l}_i-\beta \cos2\theta_2\ket{V}_s\ket{V}_i\ket{u}_i,\\
&\xrightarrow{\text{H3}}
\alpha\ket{H}_s\ket{H}_i\ket{l}_i+\beta \sin2\theta_2\ket{V}_s\ket{V}_i\ket{l}_i-\beta \cos2\theta_2 \ket{V}_s (\sin2\theta_3\ket{H}_i\ket{u}_i- \cos2\theta_3\ket{V}_i\ket{u}_i),\\
&\xrightarrow{\text{re-encode}}
\alpha\ket{000}+\beta \sin2\theta_2\ket{110}\\&-\beta \cos2\theta_2 (\sin2\theta_3\ket{101}- \cos2\theta_3\ket{111}),
\end{split}
\end{equation}
\end{widetext}
where $\theta_2$ and $\theta_3$ are the rotation angles of wave plates H2 and H3, respectively; and the polarization $H$($V$) and path mode $l$($u$) are re-encoded as $0$($1$).

Finally, we use the boxed parts in Fig.~\ref{fig:setup} to analyze the photon's state, where the blocks $(a)$, $(b)$, and $(c)$ can project a single qubit onto an arbitrary basis. Here, BD3 has two functions. On the one hand, it acts as a polarizer. With the wave plates in box $(b)$, it performs measurements on the polarization-DoF of the photon.  On the other hand, it maps the spatial modes $u$ and $l$ to the polarization modes $H$ and $V$, and coherently combines the photons in different paths, so the information of the path-DoF is analyzed via the block $(c)$. For each experimentally prepared state, we construct the detailed form of its density matrix through the standard quantum state tomography technology \cite{james2001}. All the fidelities of the states are above $0.9782\pm0.0004$, as will be shown in Table \ref{tab:exp}, which shows small deviations between the prepared states and the ideal pure states. 

{\it Pure state examples.}---Here we demonstrate the lower bounds and upper bounds of three-qubit entanglement by two different entanglement measures. 

\begin{table}[b]
    \centering
    \begin{threeparttable}
    \begin{tabular}{c||c|c|c|c|c|c}
    &$\theta_1$& $\theta_2$ & $\theta_3$ & Fidelity & Purity\\
    \hline
    \hline
    $\psi_1$ & $22.5^\circ$ & $-18^\circ$ & 0 & 0.9840 & 0.9726\\
    \hline
    $\psi_2$ & $33.75^\circ$ & 0 & 0 & 0.9829 & 0.9723\\
    \hline
    $\psi_3$ & $22.5^\circ$ & $-22.5^\circ$ & 0 & 0.9848 & 0.9746\\
    \hline
    $W$ & $27.37^\circ$ & $67.55^\circ$ & $45^\circ$ & 0.9782 & 0.9626\\
    \hline
    Bisep & $22.5^\circ$ & $45^\circ$ & NR & 0.9937  & 0.9880  \\
    \end{tabular}
    \begin{tablenotes}
    \footnotesize
    \item[*] NR: No Restriction.
    \end{tablenotes}
    \end{threeparttable}
    \caption{The fidelity and the purity for each pure state we prepared, and the corresponding settings $\{\theta_1, \theta_2, \theta_3\}$ of the wave plates \{H1, H2, H3\}. The uncertainties of the Fidelity and Purity are about $0.0004$, which are estimated from Monte Carlo simulations with 200 iterations. }
    \label{tab:exp}
\end{table}

\begin{figure*}[t]
    \centering
    \includegraphics[width=0.45\textwidth]{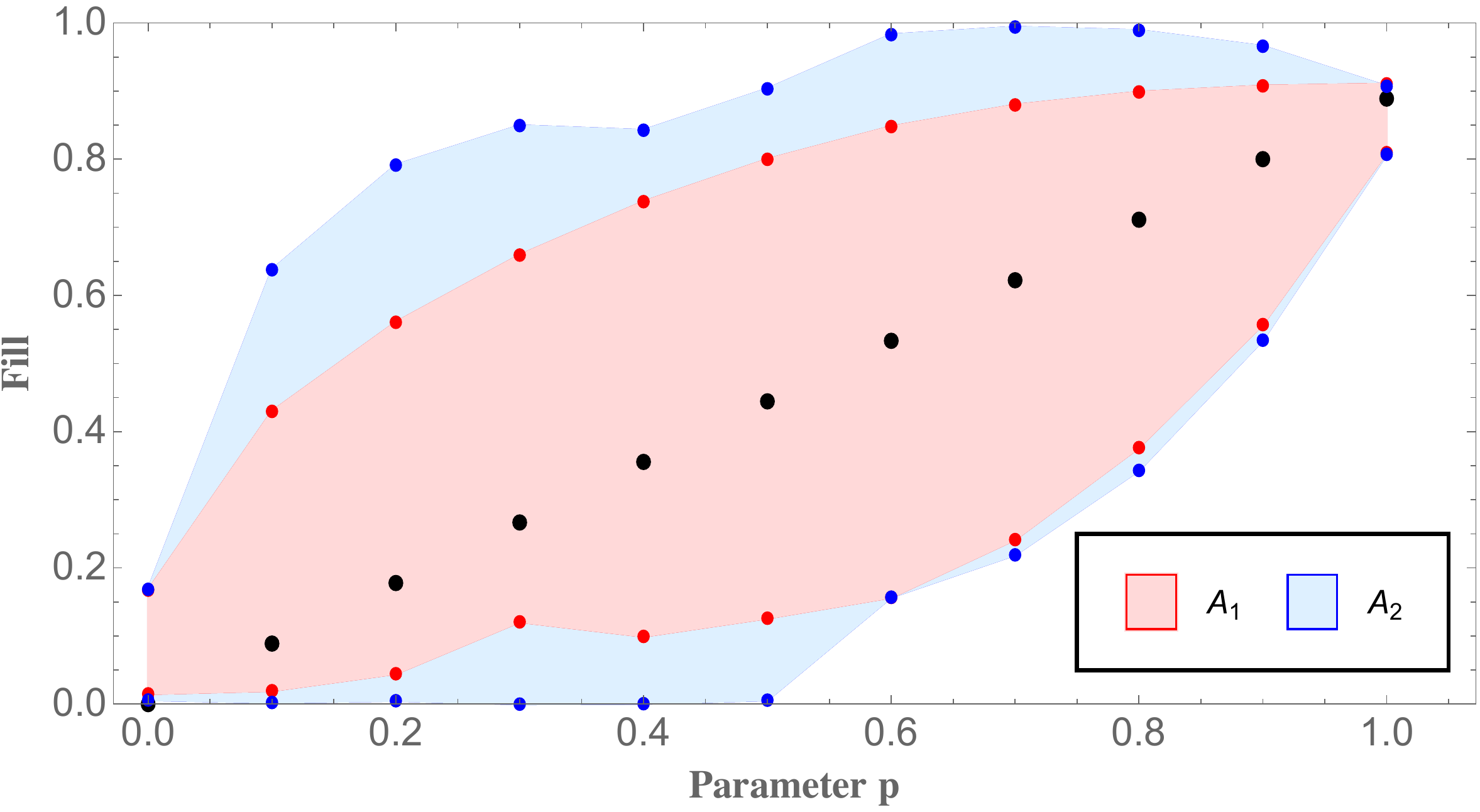}
    \includegraphics[width=0.45\textwidth]{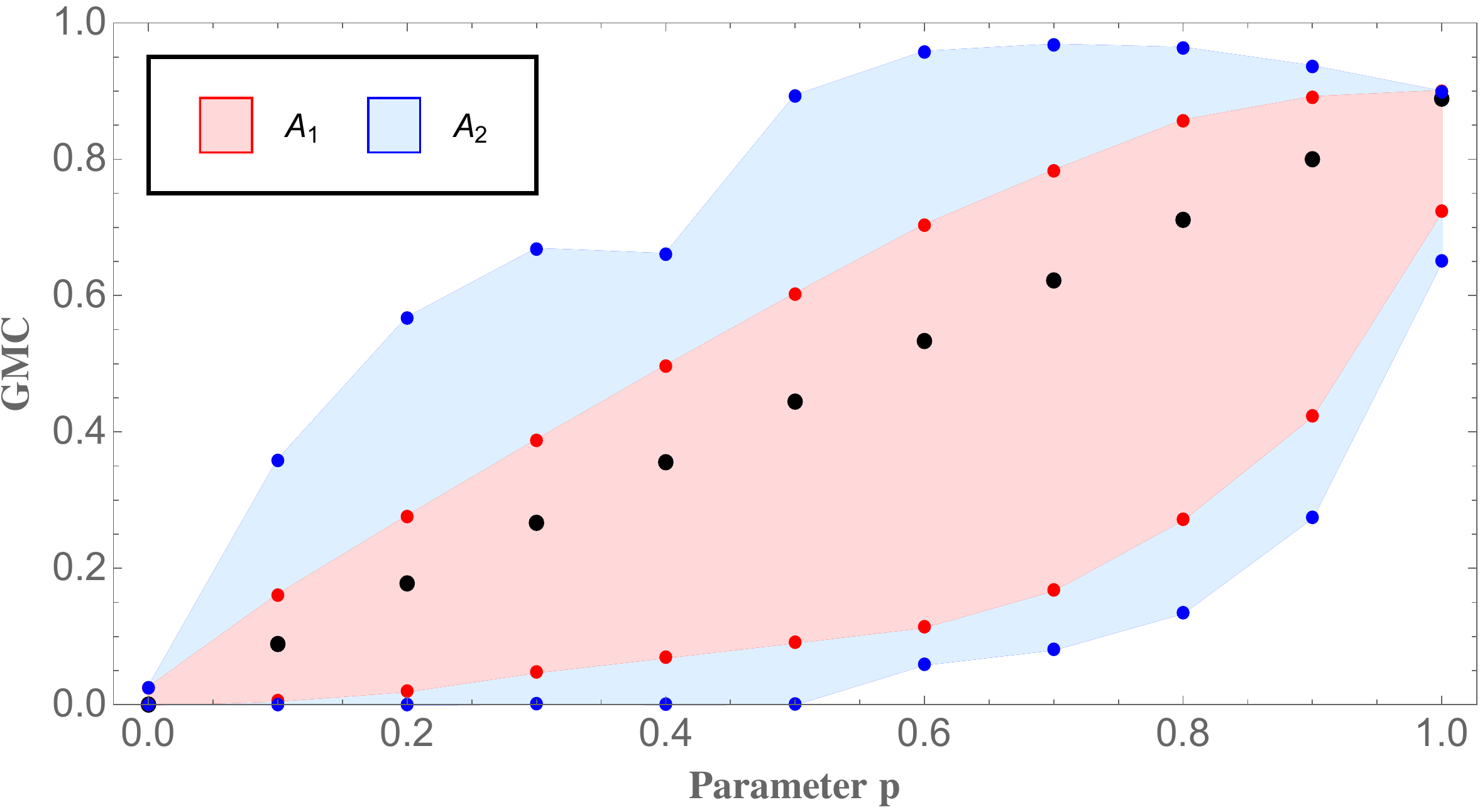}
    \caption{The entanglement bound estimations for the mixed state $(1-p)|\text{Bisep}\rangle\langle\text{Bisep}|+p|W\rangle\langle W|$ with the Fill/GMC measure. The black dots are the theoretical entanglements for the mixed state, which are close to the exact entanglement for the prepared states. The red shaded region is bounded by operator $\A_1$. The blue shaded region is bounded by operator $\A_2$. See details in the text.}
    \label{fig:fillcurve}
\end{figure*}

{\it Concurrence fill} (abbreviated as Fill) is a GME measure, the square root of the area of the concurrence triangle introduced in \cite{xie2021}. {\it Genuine multipartite concurrence} (abbreviated as GMC) is another GME measure, introduced in \cite{ma2011}, and interpreted as the length of the shortest edge for the concurrence triangle. Fill and GMC were found to be two inequivalent GME measures in the sense that they can give different entanglement rankings to the same pair of states \cite{xie2021}.

The initial definitions of the above measures are all for pure quantum states. The extensions to mixed states are given by the convex-roof construction Eq. \eqref{convexroof}. Therefore, our LB and UB estimators are able to provide estimations for these measures.

We emphasize that it is an art to choose an appropriate Hermitian operator $\A$, and thus, to choose an appropriate fiber $\A-\lambda\mathbb{1}$, so that it gives an efficiently good estimate of entanglement. There can be a trade-off between getting a better estimate and having a simpler form of $\A$. For a pure quantum state $|\psi\rangle$ prepared in an experiment, we find that the expectation value of $\A_1=x|\psi\rangle\langle\psi|$ provides good LB/UB estimates. Furthermore, only a small number of basis operators (tensor products of the Pauli operators) need to be measured. The parameter $x$ is carefully selected so that the lower(upper) bound is maximized(minimized). Note that the selection of $x$ does not increase any experimental burden.

We demonstrate results for each of the following five pure states prepared in our experiment,
\begin{equation}\label{fourstate}
    \begin{split}
        &|\psi_1\rangle=\dfrac{1}{\sqrt{2}}\left[|000\rangle+\cos(\dfrac{\pi}{5})|110\rangle+\sin(\dfrac{\pi}{5})|111\rangle\right],\\
        &|\psi_2\rangle=\cos(\dfrac{\pi}{8})|000\rangle+\sin(\dfrac{\pi}{8})|111\rangle,\\
        &|\psi_3\rangle=\dfrac{1}{\sqrt{2}}|000\rangle+\dfrac{1}{2}|110\rangle+\dfrac{1}{2}|111\rangle,\\
        &|W\rangle=\dfrac{1}{\sqrt{3}}(|100\rangle+|010\rangle+|001\rangle),\\
        &|\text{Bisep}\rangle=\dfrac{1}{\sqrt{2}}(|000\rangle+|110\rangle).
    \end{split}
\end{equation}
$|\psi_1\rangle,|\psi_2\rangle,|\psi_3\rangle$ were introduced in \cite{xie2021} with relabelings. The state $|\text{Bisep}\rangle$ is a biseparable state which contains no genuine tripartite entanglement. For these states, the settings of the half-wave plates H1$\sim$H3 in Fig.~\ref{fig:setup} are given in Table~\ref{tab:exp}. Note that, for the $W$ state, an additional half-wave plate H4 oriented at $45^{\circ}$ is needed to perform a bit flip operation on the first qubit, i.e., the polarization DoF of photon $s$.

The results are shown in Fig. \ref{fig:pureresult}. The blue bounds are given by the expectation value of the operator $\A_2=x\rho_i$, where $\rho_i$ is the full tomography of the state. If a full tomography is known, the exact entanglement $E(\rho)$ can in principle be evaluated. However, the blue bounds given by the full tomography greatly reduce the numerical burden to calculate the exact entanglement. Even so, the advantage of using the operator $\A_1$ is obvious: the experimental work is greatly reduced compared to the full tomography, while keeping relatively good entanglement estimates.

{\it Mixed-state examples.}---Now we show the results for a specific type of mixed state: the mixture of the two states $|\text{Bisep}\rangle$ and $|W\rangle$,
\begin{equation}\label{ghzw}
    \rho(p)=(1-p)|\text{Bisep}\rangle\langle\text{Bisep}|+p|W\rangle\langle W|.
\end{equation}
We prepare it by designing the setup to generate photons with state $\ket{\text{Bisep}}$ in $(1-p)T$ time and state $\ket{W}$ in $pT$ time, where $T$ is the whole time for running the experiment once. The exact entanglement values of the Fill and GMC measures for these states are the same: $(8/9)p$, as can be verified numerically using the method in \cite{rothlisberger2009}. For such a mixture, we find that the expectation value of the operator $\A_1=x|\text{Bisep}\rangle\langle\text{Bisep}|+y|W\rangle\langle W|$ can give good estimates of the entanglement. As before, the parameters $\{x,y\}$ shall be chosen carefully to optimize the performance, which does not require any extra experimental measurements.

The results for the two measures are shown in the panels of Fig. \ref{fig:fillcurve}. The blue bounds are by the expectation value of the operator $\A_2=x\rho_i$, where $x$ is again a post-selected parameter and $\rho_i$ is the full tomography of the state. The shaded regions predict the entanglement by the operators $\A_1$ and $\A_2$ respectively. The theoretical entanglement values (the black dots) are all well bounded by the estimates.

Surprisingly, the extra information provided by $A_2$ does not improve the entanglement estimation. The bounds by $\A_1$ for the Fill and the GMC measures are narrower than the ones by $\A_2$. This suggests an efficient way to estimate entanglement values in a real experiment: to measure $\langle \A_1\rangle_\rho$ is good enough. 

{\it Conclusion.}---In this work, we introduced the concept of tight lower/upper bound estimators of entanglement. These estimators are two distinct sections of the fiber bundle $\A-\lambda\mathbb{1}$, where $\A$ is a Hermitian operator and $\lambda$ is a real number specifying the fibers. With these estimators, the expectation value of an arbitrary Hermitian operator is able to universally provide lower/upper bound estimations of {\it any} mixed-state entanglement measures $E(\rho)$. Our method does not require a full tomography of the state in a real experiment, but needs only a small number of measurements. We prepared several pure and mixed states in experiments and applied our method to these states. We suggested two operators $\A_1$ and $\A_2$ to estimate the entanglement. We found that $\A_2$ works better for pure states, while $\A_1$ works better for general mixed states.

The authors thank Professor Akira Sone for valuable discussions. Financial support was provided by National Science Foundation Grants No.~PHY-1505189 and No.~PHY-1539859 (INSPIRE), the National Natural Science Foundation of China No.~11821404, No.~11734015, No.~62075208, No.~12022401, and No.~62075207, the Major Key Project of PCL, and a competitive grant from the University of Rochester.

\bibliographystyle{apsrev4-2}
\bibliography{bound}

\begin{thebibliography}{21}%
\makeatletter
\providecommand \@ifxundefined [1]{%
 \@ifx{#1\undefined}
}%
\providecommand \@ifnum [1]{%
 \ifnum #1\expandafter \@firstoftwo
 \else \expandafter \@secondoftwo
 \fi
}%
\providecommand \@ifx [1]{%
 \ifx #1\expandafter \@firstoftwo
 \else \expandafter \@secondoftwo
 \fi
}%
\providecommand \natexlab [1]{#1}%
\providecommand \enquote  [1]{``#1''}%
\providecommand \bibnamefont  [1]{#1}%
\providecommand \bibfnamefont [1]{#1}%
\providecommand \citenamefont [1]{#1}%
\providecommand \href@noop [0]{\@secondoftwo}%
\providecommand \href [0]{\begingroup \@sanitize@url \@href}%
\providecommand \@href[1]{\@@startlink{#1}\@@href}%
\providecommand \@@href[1]{\endgroup#1\@@endlink}%
\providecommand \@sanitize@url [0]{\catcode `\\12\catcode `\$12\catcode
  `\&12\catcode `\#12\catcode `\^12\catcode `\_12\catcode `\%12\relax}%
\providecommand \@@startlink[1]{}%
\providecommand \@@endlink[0]{}%
\providecommand \url  [0]{\begingroup\@sanitize@url \@url }%
\providecommand \@url [1]{\endgroup\@href {#1}{\urlprefix }}%
\providecommand \urlprefix  [0]{URL }%
\providecommand \Eprint [0]{\href }%
\providecommand \doibase [0]{https://doi.org/}%
\providecommand \selectlanguage [0]{\@gobble}%
\providecommand \bibinfo  [0]{\@secondoftwo}%
\providecommand \bibfield  [0]{\@secondoftwo}%
\providecommand \translation [1]{[#1]}%
\providecommand \BibitemOpen [0]{}%
\providecommand \bibitemStop [0]{}%
\providecommand \bibitemNoStop [0]{.\EOS\space}%
\providecommand \EOS [0]{\spacefactor3000\relax}%
\providecommand \BibitemShut  [1]{\csname bibitem#1\endcsname}%
\let\auto@bib@innerbib\@empty
\bibitem [{\citenamefont {Ekert}(1991)}]{ekert1991}%
  \BibitemOpen
  \bibfield  {author} {\bibinfo {author} {\bibfnamefont {A.~K.}\ \bibnamefont
  {Ekert}},\ }\href@noop {} {\bibfield  {journal} {\bibinfo  {journal} {Phys.
  Rev. Lett.}\ }\textbf {\bibinfo {volume} {67}},\ \bibinfo {pages} {661}
  (\bibinfo {year} {1991})}\BibitemShut {NoStop}%
\bibitem [{\citenamefont {Bennett}\ and\ \citenamefont
  {Wiesner}(1992)}]{bennett1992}%
  \BibitemOpen
  \bibfield  {author} {\bibinfo {author} {\bibfnamefont {C.~H.}\ \bibnamefont
  {Bennett}}\ and\ \bibinfo {author} {\bibfnamefont {S.~J.}\ \bibnamefont
  {Wiesner}},\ }\href@noop {} {\bibfield  {journal} {\bibinfo  {journal} {Phys.
  Rev. Lett.}\ }\textbf {\bibinfo {volume} {69}},\ \bibinfo {pages} {2881}
  (\bibinfo {year} {1992})}\BibitemShut {NoStop}%
\bibitem [{\citenamefont {Bennett}\ \emph {et~al.}(1993)\citenamefont
  {Bennett}, \citenamefont {Brassard}, \citenamefont {Cr\'epeau}, \citenamefont
  {Jozsa}, \citenamefont {Peres},\ and\ \citenamefont
  {Wootters}}]{bennett1993}%
  \BibitemOpen
  \bibfield  {author} {\bibinfo {author} {\bibfnamefont {C.~H.}\ \bibnamefont
  {Bennett}}, \bibinfo {author} {\bibfnamefont {G.}~\bibnamefont {Brassard}},
  \bibinfo {author} {\bibfnamefont {C.}~\bibnamefont {Cr\'epeau}}, \bibinfo
  {author} {\bibfnamefont {R.}~\bibnamefont {Jozsa}}, \bibinfo {author}
  {\bibfnamefont {A.}~\bibnamefont {Peres}},\ and\ \bibinfo {author}
  {\bibfnamefont {W.~K.}\ \bibnamefont {Wootters}},\ }\href@noop {} {\bibfield
  {journal} {\bibinfo  {journal} {Phys. Rev. Lett.}\ }\textbf {\bibinfo
  {volume} {70}},\ \bibinfo {pages} {1895} (\bibinfo {year}
  {1993})}\BibitemShut {NoStop}%
\bibitem [{\citenamefont {Bennett}\ \emph {et~al.}(1996)\citenamefont
  {Bennett}, \citenamefont {Brassard}, \citenamefont {Popescu}, \citenamefont
  {Schumacher}, \citenamefont {Smolin},\ and\ \citenamefont
  {Wootters}}]{bennett1996}%
  \BibitemOpen
  \bibfield  {author} {\bibinfo {author} {\bibfnamefont {C.~H.}\ \bibnamefont
  {Bennett}}, \bibinfo {author} {\bibfnamefont {G.}~\bibnamefont {Brassard}},
  \bibinfo {author} {\bibfnamefont {S.}~\bibnamefont {Popescu}}, \bibinfo
  {author} {\bibfnamefont {B.}~\bibnamefont {Schumacher}}, \bibinfo {author}
  {\bibfnamefont {J.~A.}\ \bibnamefont {Smolin}},\ and\ \bibinfo {author}
  {\bibfnamefont {W.~K.}\ \bibnamefont {Wootters}},\ }\href@noop {} {\bibfield
  {journal} {\bibinfo  {journal} {Phys. Rev. Lett.}\ }\textbf {\bibinfo
  {volume} {76}},\ \bibinfo {pages} {722} (\bibinfo {year} {1996})}\BibitemShut
  {NoStop}%
\bibitem [{\citenamefont {Von~Neumann}(1955)}]{von1955}%
  \BibitemOpen
  \bibfield  {author} {\bibinfo {author} {\bibfnamefont {J.}~\bibnamefont
  {Von~Neumann}},\ }\href@noop {} {\emph {\bibinfo {title} {Mathematical
  Foundations of Quantum Mechanics}}}\ (\bibinfo  {publisher} {Princeton
  University Press},\ \bibinfo {year} {1955})\BibitemShut {NoStop}%
\bibitem [{\citenamefont {Ma}\ \emph {et~al.}(2011)\citenamefont {Ma},
  \citenamefont {Chen}, \citenamefont {Chen}, \citenamefont {Spengler},
  \citenamefont {Gabriel},\ and\ \citenamefont {Huber}}]{ma2011}%
  \BibitemOpen
  \bibfield  {author} {\bibinfo {author} {\bibfnamefont {Z.-H.}\ \bibnamefont
  {Ma}}, \bibinfo {author} {\bibfnamefont {Z.-H.}\ \bibnamefont {Chen}},
  \bibinfo {author} {\bibfnamefont {J.-L.}\ \bibnamefont {Chen}}, \bibinfo
  {author} {\bibfnamefont {C.}~\bibnamefont {Spengler}}, \bibinfo {author}
  {\bibfnamefont {A.}~\bibnamefont {Gabriel}},\ and\ \bibinfo {author}
  {\bibfnamefont {M.}~\bibnamefont {Huber}},\ }\href@noop {} {\bibfield
  {journal} {\bibinfo  {journal} {Phys. Rev. A}\ }\textbf {\bibinfo {volume}
  {83}},\ \bibinfo {pages} {062325} (\bibinfo {year} {2011})}\BibitemShut
  {NoStop}%
\bibitem [{\citenamefont {Uhlmann}(1998)}]{uhlmann1998}%
  \BibitemOpen
  \bibfield  {author} {\bibinfo {author} {\bibfnamefont {A.}~\bibnamefont
  {Uhlmann}},\ }\href@noop {} {\bibfield  {journal} {\bibinfo  {journal} {Open
  Systems \& Information Dynamics}\ }\textbf {\bibinfo {volume} {5}},\ \bibinfo
  {pages} {209} (\bibinfo {year} {1998})}\BibitemShut {NoStop}%
\bibitem [{\citenamefont {Hill}\ and\ \citenamefont
  {Wootters}(1997)}]{hill1997}%
  \BibitemOpen
  \bibfield  {author} {\bibinfo {author} {\bibfnamefont {S.~A.}\ \bibnamefont
  {Hill}}\ and\ \bibinfo {author} {\bibfnamefont {W.~K.}\ \bibnamefont
  {Wootters}},\ }\href@noop {} {\bibfield  {journal} {\bibinfo  {journal}
  {Phys. Rev. Lett.}\ }\textbf {\bibinfo {volume} {78}},\ \bibinfo {pages}
  {5022} (\bibinfo {year} {1997})}\BibitemShut {NoStop}%
\bibitem [{\citenamefont {R{\"o}thlisberger}\ \emph {et~al.}(2009)\citenamefont
  {R{\"o}thlisberger}, \citenamefont {Lehmann},\ and\ \citenamefont
  {Loss}}]{rothlisberger2009}%
  \BibitemOpen
  \bibfield  {author} {\bibinfo {author} {\bibfnamefont {B.}~\bibnamefont
  {R{\"o}thlisberger}}, \bibinfo {author} {\bibfnamefont {J.}~\bibnamefont
  {Lehmann}},\ and\ \bibinfo {author} {\bibfnamefont {D.}~\bibnamefont
  {Loss}},\ }\href@noop {} {\bibfield  {journal} {\bibinfo  {journal} {Physical
  Review A}\ }\textbf {\bibinfo {volume} {80}},\ \bibinfo {pages} {042301}
  (\bibinfo {year} {2009})}\BibitemShut {NoStop}%
\bibitem [{\citenamefont {Audenaert}\ and\ \citenamefont
  {Plenio}(2006)}]{audenaert2006}%
  \BibitemOpen
  \bibfield  {author} {\bibinfo {author} {\bibfnamefont {K.~M.}\ \bibnamefont
  {Audenaert}}\ and\ \bibinfo {author} {\bibfnamefont {M.~B.}\ \bibnamefont
  {Plenio}},\ }\href@noop {} {\bibfield  {journal} {\bibinfo  {journal} {New
  Journal of Physics}\ }\textbf {\bibinfo {volume} {8}},\ \bibinfo {pages}
  {266} (\bibinfo {year} {2006})}\BibitemShut {NoStop}%
\bibitem [{\citenamefont {G\"uhne}\ \emph {et~al.}(2007)\citenamefont
  {G\"uhne}, \citenamefont {Reimpell},\ and\ \citenamefont
  {Werner}}]{guhne2007}%
  \BibitemOpen
  \bibfield  {author} {\bibinfo {author} {\bibfnamefont {O.}~\bibnamefont
  {G\"uhne}}, \bibinfo {author} {\bibfnamefont {M.}~\bibnamefont {Reimpell}},\
  and\ \bibinfo {author} {\bibfnamefont {R.~F.}\ \bibnamefont {Werner}},\
  }\href@noop {} {\bibfield  {journal} {\bibinfo  {journal} {Phys. Rev. Lett.}\
  }\textbf {\bibinfo {volume} {98}},\ \bibinfo {pages} {110502} (\bibinfo
  {year} {2007})}\BibitemShut {NoStop}%
\bibitem [{\citenamefont {Eisert}\ \emph {et~al.}(2007)\citenamefont {Eisert},
  \citenamefont {Brand\~{a}o},\ and\ \citenamefont {Audenaert}}]{eisert2007}%
  \BibitemOpen
  \bibfield  {author} {\bibinfo {author} {\bibfnamefont {J.}~\bibnamefont
  {Eisert}}, \bibinfo {author} {\bibfnamefont {F.~G.}\ \bibnamefont
  {Brand\~{a}o}},\ and\ \bibinfo {author} {\bibfnamefont {K.~M.}\ \bibnamefont
  {Audenaert}},\ }\href@noop {} {\bibfield  {journal} {\bibinfo  {journal} {New
  Journal of Physics}\ }\textbf {\bibinfo {volume} {9}},\ \bibinfo {pages} {46}
  (\bibinfo {year} {2007})}\BibitemShut {NoStop}%
\bibitem [{\citenamefont {Yu}\ and\ \citenamefont {Eberly}(2009)}]{yu2009}%
  \BibitemOpen
  \bibfield  {author} {\bibinfo {author} {\bibfnamefont {T.}~\bibnamefont
  {Yu}}\ and\ \bibinfo {author} {\bibfnamefont {J.~H.}\ \bibnamefont
  {Eberly}},\ }\href@noop {} {\bibfield  {journal} {\bibinfo  {journal}
  {Science}\ }\textbf {\bibinfo {volume} {323}},\ \bibinfo {pages} {598}
  (\bibinfo {year} {2009})}\BibitemShut {NoStop}%
\bibitem [{\citenamefont {Zhu}\ and\ \citenamefont {Fei}(2012)}]{zhu2012}%
  \BibitemOpen
  \bibfield  {author} {\bibinfo {author} {\bibfnamefont {X.-N.}\ \bibnamefont
  {Zhu}}\ and\ \bibinfo {author} {\bibfnamefont {S.-M.}\ \bibnamefont {Fei}},\
  }\href@noop {} {\bibfield  {journal} {\bibinfo  {journal} {Phys. Rev. A}\
  }\textbf {\bibinfo {volume} {86}},\ \bibinfo {pages} {054301} (\bibinfo
  {year} {2012})}\BibitemShut {NoStop}%
\bibitem [{\citenamefont {Song}\ \emph {et~al.}(2016)\citenamefont {Song},
  \citenamefont {Chen},\ and\ \citenamefont {Cao}}]{song2016}%
  \BibitemOpen
  \bibfield  {author} {\bibinfo {author} {\bibfnamefont {W.}~\bibnamefont
  {Song}}, \bibinfo {author} {\bibfnamefont {L.}~\bibnamefont {Chen}},\ and\
  \bibinfo {author} {\bibfnamefont {Z.-L.}\ \bibnamefont {Cao}},\ }\href@noop
  {} {\bibfield  {journal} {\bibinfo  {journal} {Scientific Reports}\ }\textbf
  {\bibinfo {volume} {6}},\ \bibinfo {pages} {23} (\bibinfo {year}
  {2016})}\BibitemShut {NoStop}%
\bibitem [{\citenamefont {Ryu}\ \emph {et~al.}(2012)\citenamefont {Ryu},
  \citenamefont {Lee},\ and\ \citenamefont {Sim}}]{ryu2012}%
  \BibitemOpen
  \bibfield  {author} {\bibinfo {author} {\bibfnamefont {S.}~\bibnamefont
  {Ryu}}, \bibinfo {author} {\bibfnamefont {S.-S.~B.}\ \bibnamefont {Lee}},\
  and\ \bibinfo {author} {\bibfnamefont {H.-S.}\ \bibnamefont {Sim}},\
  }\href@noop {} {\bibfield  {journal} {\bibinfo  {journal} {Phys. Rev. A}\
  }\textbf {\bibinfo {volume} {86}},\ \bibinfo {pages} {042324} (\bibinfo
  {year} {2012})}\BibitemShut {NoStop}%
\bibitem [{\citenamefont {Far{\'\i}as}\ \emph {et~al.}(2012)\citenamefont
  {Far{\'\i}as}, \citenamefont {Aguilar}, \citenamefont
  {Vald{\'e}s-Hern{\'a}ndez}, \citenamefont {Ribeiro}, \citenamefont
  {Davidovich},\ and\ \citenamefont {Walborn}}]{Osetup2012}%
  \BibitemOpen
  \bibfield  {author} {\bibinfo {author} {\bibfnamefont {O.~J.}\ \bibnamefont
  {Far{\'\i}as}}, \bibinfo {author} {\bibfnamefont {G.}~\bibnamefont
  {Aguilar}}, \bibinfo {author} {\bibfnamefont {A.}~\bibnamefont
  {Vald{\'e}s-Hern{\'a}ndez}}, \bibinfo {author} {\bibfnamefont {P.~S.}\
  \bibnamefont {Ribeiro}}, \bibinfo {author} {\bibfnamefont {L.}~\bibnamefont
  {Davidovich}},\ and\ \bibinfo {author} {\bibfnamefont {S.~P.}\ \bibnamefont
  {Walborn}},\ }\href@noop {} {\bibfield  {journal} {\bibinfo  {journal} {Phys.
  Rev. Lett.}\ }\textbf {\bibinfo {volume} {109}},\ \bibinfo {pages} {150403}
  (\bibinfo {year} {2012})}\BibitemShut {NoStop}%
\bibitem [{\citenamefont {Aguilar}\ \emph {et~al.}(2015)\citenamefont
  {Aguilar}, \citenamefont {Walborn}, \citenamefont {Ribeiro},\ and\
  \citenamefont {C{\'e}leri}}]{Osetup2015}%
  \BibitemOpen
  \bibfield  {author} {\bibinfo {author} {\bibfnamefont {G.~H.}\ \bibnamefont
  {Aguilar}}, \bibinfo {author} {\bibfnamefont {S.~P.}\ \bibnamefont
  {Walborn}}, \bibinfo {author} {\bibfnamefont {P.~S.}\ \bibnamefont
  {Ribeiro}},\ and\ \bibinfo {author} {\bibfnamefont {L.~C.}\ \bibnamefont
  {C{\'e}leri}},\ }\href@noop {} {\bibfield  {journal} {\bibinfo  {journal}
  {Physical Review X}\ }\textbf {\bibinfo {volume} {5}},\ \bibinfo {pages}
  {031042} (\bibinfo {year} {2015})}\BibitemShut {NoStop}%
\bibitem [{\citenamefont {Zhang}\ \emph {et~al.}(2021)\citenamefont {Zhang},
  \citenamefont {Sun}, \citenamefont {Fang}, \citenamefont {Zhang},
  \citenamefont {Yuan},\ and\ \citenamefont {Lu}}]{Osetup2021}%
  \BibitemOpen
  \bibfield  {author} {\bibinfo {author} {\bibfnamefont {T.}~\bibnamefont
  {Zhang}}, \bibinfo {author} {\bibfnamefont {J.}~\bibnamefont {Sun}}, \bibinfo
  {author} {\bibfnamefont {X.-X.}\ \bibnamefont {Fang}}, \bibinfo {author}
  {\bibfnamefont {X.-M.}\ \bibnamefont {Zhang}}, \bibinfo {author}
  {\bibfnamefont {X.}~\bibnamefont {Yuan}},\ and\ \bibinfo {author}
  {\bibfnamefont {H.}~\bibnamefont {Lu}},\ }\href@noop {} {\bibfield  {journal}
  {\bibinfo  {journal} {Phys. Rev. Lett.}\ }\textbf {\bibinfo {volume} {127}},\
  \bibinfo {pages} {200501} (\bibinfo {year} {2021})}\BibitemShut {NoStop}%
\bibitem [{\citenamefont {James}\ \emph {et~al.}(2001)\citenamefont {James},
  \citenamefont {Kwiat}, \citenamefont {Munro},\ and\ \citenamefont
  {White}}]{james2001}%
  \BibitemOpen
  \bibfield  {author} {\bibinfo {author} {\bibfnamefont {D.~F.~V.}\
  \bibnamefont {James}}, \bibinfo {author} {\bibfnamefont {P.~G.}\ \bibnamefont
  {Kwiat}}, \bibinfo {author} {\bibfnamefont {W.~J.}\ \bibnamefont {Munro}},\
  and\ \bibinfo {author} {\bibfnamefont {A.~G.}\ \bibnamefont {White}},\
  }\href@noop {} {\bibfield  {journal} {\bibinfo  {journal} {Phys. Rev. A}\
  }\textbf {\bibinfo {volume} {64}},\ \bibinfo {pages} {052312} (\bibinfo
  {year} {2001})}\BibitemShut {NoStop}%
\bibitem [{\citenamefont {Xie}\ and\ \citenamefont {Eberly}(2021)}]{xie2021}%
  \BibitemOpen
  \bibfield  {author} {\bibinfo {author} {\bibfnamefont {S.}~\bibnamefont
  {Xie}}\ and\ \bibinfo {author} {\bibfnamefont {J.~H.}\ \bibnamefont
  {Eberly}},\ }\href@noop {} {\bibfield  {journal} {\bibinfo  {journal} {Phys.
  Rev. Lett.}\ }\textbf {\bibinfo {volume} {127}},\ \bibinfo {pages} {040403}
  (\bibinfo {year} {2021})}\BibitemShut {NoStop}%
\end{thebibliography}%

\end{document}